\date{April 4,  2002}

\documentclass[reqno]{amsart}
\usepackage{amsmath,amsfonts,amssymb,amsthm}
\input epsf

\numberwithin{equation}{section}
\newtheorem{thm}{Theorem}
\newtheorem{lemma}{Lemma}[section]

\newtheorem{prop}{Proposition}
\newcommand{\remark}{{\it Remark: \,\,}}

\newcommand{\abs}[1]{ \left | #1 \right |}
\newcommand{\dist}{{\, \rm dist}}

\newcommand{\com}[2]{ \left [ #1 \, , #2 \right ] }

\newcommand{\Z}{\mathbb Z}
\newcommand{\Hi}{\mathcal H}

\newcommand{\sprod}[2]{\mbox{$\langle #1,#2 \rangle$}}      
\newcommand{\e}{\mathrm e}
\newcommand{\im}{\mathrm i}

\newcommand{\1}{{\mathbf 1}}
\newcommand{\R}{\mathbb R}
\newcommand{\N}{\mathbb N}

\newcommand{\ph}{\varphi}

\newcommand{\Sch}{\mathcal S}

\newcommand{\di}{\mathrm d}

\newcommand{\ax}{\langle x \rangle}
\newcommand{\axvec}{\langle \bf{x} \rangle}
\newcommand{\bx}{\langle x_1 \rangle}
\newcommand{\cx}{\langle x_2 \rangle}
\newcommand{\axi}{\langle x_i \rangle}
\newcommand{\axone}{\langle x_1 \rangle}
\newcommand{\axtwo}{\langle x_2 \rangle}

\newcommand{\first}{$1^{\rm st}$ }
\newcommand{\second}{$2^{\rm nd}$ }
\DeclareMathOperator{\Tr}{Tr}

\newcounter{list}

\theoremstyle{definition}
\newtheorem{defn}{Definition}

\begin{document}
\title[Adiabatic charge transport and the Kubo formula]{Adiabatic charge
transport and  the Kubo formula for Landau Type Hamiltonians}
\author[A. Elgart]{Alexander Elgart}
\author[B. Schlein]{Benjamin Schlein}
\address{Courant Institute, 251 Mercer St., New York, NY 10012, USA.}
\email{elgart@cims.nyu.edu, schlein@cims.nyu.edu} 
\begin{abstract}
The adiabatic charge transport is investigated in a two dimensional Landau model 
perturbed by a bounded potential at zero temperature. We show that 
if the Fermi level lies in a spectral gap then in the adiabatic limit 
the accumulated excess Hall transport is given by the linear
response Kubo-\v{S}treda formula. The proof relies on the expansion 
of Nenciu, some generalized phase space estimates, and 
a bound on the speed of propagation.
\end{abstract}

\date{04/15/2003}

\maketitle

\section{Introduction}
In this work we prove the validity of the linear response theory in 
the context of the Landau type model of non-interacting electrons. 
The framework of the linear response was used extensively to explain
 the quantization of conductance in the Quantum Hall Effect (QHE) 
\cite{Laugh81,Th81,ASY}, first observed in the celebrated experiment by von
Klitzing et. al. \cite{KDP}. Most of the results derived in this paper 
can be proved for  more general families of magnetic 
Schr\"odinger operators in $\R^2$ than the Landau 
model, but the latter is {\em primus inter pares} for us due to its 
accessibility. It is also worthwhile to note  that as far as we know, the 
Landau model is the only model 
in $\R^2$ for which the (quantized) conductivity was actually computed (see,
for example \cite{AS2} and references therein).

The model we consider here is described by a one-particle 
Hamiltonian on $L^2(\R^2)$:
 $$H_{\lambda} \ = \ (p_1-\frac{B}{2}x_2)^2 \ + \  (p_2+\frac{B}{2}x_1)^2\ + \ 
\lambda V\,,$$ where the potential $ V$ is smooth
and relatively weak: $\|V\|_{n,\infty}\le C_n$ for some integer $n$ large enough
($n=5$ will do), and $\| V\|_\infty < B/ \lambda$. The operator $H_0$ has been 
introduced by Lev 
Landau, and has a number of nice features. In particular, its spectrum consists
of the odd multiples of the strength of the magnetic field $B$. The 
corresponding eigenvalues are infinitely degenerate. Consequently, 
the spectrum of the Hamiltonian $H_\lambda$, under the above assumptions,
contains a
(infinite) sequence of bands, separated from each other by finite
gaps. In order to investigate the transport properties of the above system, 
we consider the transverse current
induced by a time-dependent potential gradient. The full
Hamiltonian  is
\begin{equation}\label{eq:11}
H_{\lambda}(t) \ = \ H_{\lambda} \ + \ \frac{1}{\tau} \,
g(t/\tau)   \, \Lambda_1 \; ,
\end{equation}
where $\dot{g}(\cdot)$ is a smooth function supported in $(0,1)$ (without loss
of generality we will assume that $g$ vanishes for negative values of the 
argument). The variable $t$ here stands for the time, and the large parameter $\tau$ is a convenient tool to 
control the rate at which the system changes. We study here the evolution up 
to time $\tau$. The 
potential $\Lambda_1$ is taken here to be of the form of a smooth
step function depending only on the $x_{1}$ component of the
position, which is $0$ on $[-\infty,-m]$ and $1$ on $[m,+\infty]$
for some $m>0$. The coefficient $1/\tau$ in front of the time dependent 
potential tells us that the external field is weak. In fact, we will
 be interested in the limiting behavior $\tau\rightarrow\infty$.

 A well known weakness of standard linear response theory, stressed by 
van Kampen \cite{VK}, is that it takes the limit of weak field first and only 
then the thermodynamic limit. The correct order is, of course, the reverse. 
Theorem \ref{thm} below is free of this criticism in that one starts with 
a system that has infinite extent. The linear response limit is then realized 
by our adiabatic limit $\tau\rightarrow\infty$.  

  
 We will compute the induced current across the line $x_{2}=0$,
in the state $\varrho_\tau(t) $ which evolves from the zero
temperature state $P_\lambda \ = \ \chi(H_\lambda < E_f)$, with
$E_f$ the Fermi-energy. Our results here deal with values of
the Fermi energy which lie in a gap of the spectrum of
$H_\lambda$.

The expectation value of the excess current is given by
\begin{equation}\label{eq:deltaJ}
 \mathcal J_\tau (t) \ = \ - \im \, \Tr \left ( \varrho_\tau(t) -  P_\lambda
\right ) \, \com{H_\lambda}{\Lambda_2} \; ,
\end{equation}
where $\Lambda_2$ is the characteristic function of the upper half
plane ($x_2\ge0$). The subtraction eliminates the contribution 
present even without the external field (persistent current). 

We will prove that, if the
Fermi energy falls in a spectral gap, then the combination
appearing in Eq.~(\ref{eq:deltaJ}) forms a trace-class operator
(which implies that the r.h.s. of Eq. (\ref{eq:deltaJ}) is well defined). 
Moreover, the main result of this paper will be  
\begin{equation}
\lim_{\tau \rightarrow \infty} \tau \mathcal J_\tau (t)\  = \ -\im  g(t/\tau)
 \Tr P_\lambda
   \com{ \,  \com{P_\lambda}{\Lambda_1}}{\com{P_\lambda}{\Lambda_2}
   \, } P_\lambda \;  .
\label{eq:k1}
\end{equation}
 This agrees with the
Kubo-\v{S}treda formula for the bulk Hall conductance \cite{kubo2,streda,Kubo}, 
confirming the validity of the linear response calculation in this context. The
equality of the bulk and the edge conductance was rigorously established in
\cite{KRS,Gr}.

Moreover it turns out that  the  quantity
\begin{equation}\label{eq:Klambda}
K_\lambda:=\Tr P_\lambda
   \com{ \,  \com{P_\lambda}{\Lambda_1}}{\com{P_\lambda}{\Lambda_2}
   \, } P_\lambda
\end{equation} 
defined in Eq. (\ref{eq:k1}) remains invariant under variation of 
$\lambda$, and in particular
\begin{equation}
K_\lambda  =  K_0\;  , \label{eq:kl}
\end{equation}
with $K_0$ being the Hall conductance at energy $E_F$ for the
Landau model. The latter was studied extensively by a number of
authors \cite{NTW,AS2,BES}, and can be computed explicitly. Whenever
the Fermi energy $E_F$ lies in the gap between the $j$th and
$j+1$th Landau level, the corresponding value of $K_0$ is equal to
$j$.

A key role in the proof is played by an
asymptotic expansion developed by Nenciu \cite{nenciu} for an
adiabatically evolved projection operator, and by the propagation 
estimate of the evolution operator, generated by the Hamiltonian
$H_{\lambda}(t)$. The physical intuition behind the proof will be presented in
details in Section \ref{sec:Nenciu}

Our results are essentially parallel to the ones derived for lattice models in 
\cite{AES}. The important difference however lies in the absence of the 
ultraviolet cutoff in the continuous case, which affects both the trace class 
properties of the relevant operators and the propagation estimates.  

There is a large amount of literature related to the different features of 
Landau type systems in both physical (e.g. \cite{isi, itzy}) 
and mathematical literature, see for example \cite{wang} 
and references therein. 
This work is not the first one where the linear response theory is discussed
in this context. We are aware of two works on this subject, 
\cite{AS85,kunz}. In \cite{AS85} the validity of the linear response theory 
is proven for finite dimensional spectral projection of magnetic Hamiltonian 
in the torus geometry. The method derived there enables one to compute the charge 
transport in the system, where ours also gives a pointwise value for a current. 
In \cite{kunz} the author derives the Hall conductance of the infinite sample 
under assumption of the validity of the linear response theory following Kubo
\cite{kubo2}, and also proves the stability of the Hall conductance 
assuming that the strength of the (random) potential is weak enough and
that states near the edges of the Landau bands are sufficiently localized. We
don't use the latter condition. 

The flow of the paper is organized as follows: in Section
\ref{sec:main} we state our main results, in Section
\ref{sec:Nenciu} we outline the Nenciu expansion and present the
basic propositions needed for our analysis: Generalized space-momentum
inequalities, finite speed of propagation for certain
initial data, and trace class estimates. Equipped with these tools, we prove
our main result, Theorem \ref{thm}. In Sections \ref{sec:tr} and \ref{sec:propag}
we outline the proofs of the aforementioned bounds. 

\section{Statement of the main result}\label{sec:main}

We require certain technical assumptions on the Hamiltonian and
the perturbing potential.

The external potential considered here will be in the form of a
{\it switch function}, $\Lambda_1$, in the terminology of
ref.~(\cite{AS85}).
\begin{defn}
A {\em switch function in the $j^{\rm th}$-direction} with $j=1$
or $2$ is a smooth function $\Lambda_j:\R^2 \rightarrow [0,1]$
which depends only on the variable $x_j$ and satisfies
\begin{equation}
\begin{cases}
\Lambda_j \ = 0 \ & x_j < -m \\ 1 - \Lambda_j \ =0 \  & x_j > m
\end{cases} \; ,
\label{eq:switch}
\end{equation}
for some $m > 0$.
\end{defn}
\noindent \remark The condition above could be replaced with
sufficiently fast power decay of the derivative of $\Lambda_j$ at
infinities without affecting the results which appear below.

The following theorem consists of three parts. The first statement shows that
the current defined in (\ref{eq:deltaJ}) is well defined, the second proves the
convergence of the charge, in the adiabatic limit, to the value given by the
Kubo formula, and the third one deals with the stability of the limiting value
of the charge under changes of $\lambda$.  

\begin{thm}\label{thm}
Let $H_\lambda$ be as above, with $a:=\lambda\, \|V\|_\infty<B$.
Assume in addition that the Sobolev norm of the potential,
$\|V\|_{N,\infty}$ is bounded for $N$ large enough ($N=6$ will do). Let
$\varrho_\tau$ be the solution to the evolution equation, for $
t\in [0, \tau]$:
\begin{equation}\label{eq:thmheis}
\begin{cases}
\im \dot \varrho_\tau(t) \ = \ \com{H_\lambda + \frac{1}{\tau} g(t
/ \tau) \, \Lambda_1}{ \varrho_\tau(t) } \\ \, \varrho_\tau(0) \ =
\ \chi(H_\lambda < E_F) \ =: \ P_\lambda
\end{cases} \; ,
\end{equation}
with $\Lambda_1$ a switch function in the \first direction,
$(2j-1)B + a < E_F <(2j+1) B - a$ for some integer $j$, and $\dot{g} \in  C^k 
([0,1])$, $g(0)=0$ ($k=4$ will be enough). 
Then for any switch function in the \second direction,
$\Lambda_2$,
\begin{list}{(\arabic{list})}{\usecounter{list}
\setlength{\leftmargin=25pt} \setlength{\rightmargin=20pt}
\setlength{\labelsep=10pt} \setlength{\itemindent=10pt}
\setlength{\itemsep=5pt}}
\item{}[Well-posedness] The observable whose trace is the induced Hall current,
\begin{equation}
\widehat {\mathcal J}_\tau(t) := -\im(\varrho_\tau(t) - P_0) \com{
H_\lambda} {\Lambda_2} \; ,
\end{equation}
is trace class for all $\tau \ge 0$ and $t \in [0,\tau]$.
\item{}[Adiabatic limit and the Kubo formula] 
For every $\tau > 0$ and $t \in [0,\tau]$ we have 
\begin{equation}\label{eq:current}
\abs{ \Tr  \widehat {\mathcal J}_\tau(t) \ + \ \frac{1 }{\tau} 
\im g(t/\tau) K_\lambda }  \ \le \ \frac{C}{\tau^2}\, ,
\end{equation}
where $K_{\lambda}$ is defined in Eq. (\ref{eq:Klambda}). 
\item{}[Stability] \label{thm:stab} The value of $K_\lambda$ is independent of the 
value of $\lambda$
whenever $a< B$. If $E_F$ satisfies the conditions above, then $K_\lambda=j$.
\end{list}
\end{thm}

\noindent {\it Remarks}
\begin{list}{(\arabic{list})}{\usecounter{list}
\setlength{\leftmargin=25pt} \setlength{\rightmargin=20pt}
\setlength{\labelsep=10pt} \setlength{\itemindent=10pt}
\setlength{\itemsep=5pt}}
\item  Eq.~\eqref{eq:current} implies that the limiting value of the 
charge is given by
\begin{equation}
\lim_{\tau \rightarrow \infty} \int_0^\tau \Tr \widehat {\mathcal
J}_\tau(t) \, \di t \  = \ -\im \, 
\left(\int_0^1 g(u) du \right)  \Tr P_\lambda
   \com{ \,  \com{P_\lambda}{\Lambda_1}}{\com{P_\lambda}
   {\Lambda_2} \, } P_\lambda \; ,
\end{equation}
known as the Kubo formula.

\item In order to prove \eqref{eq:current} we will expand $\widehat{\mathcal J}
(\tau s)$ in an asymptotic series $\sum_{j=1}^\infty \frac{1}{\tau^j} N_j(s)$.
Assuming that $g$ and its first $k+2$ derivatives vanish
at $0$, we will then prove that
\begin{equation}\label{eq:asymptotic}
\sup_{s \in [0,1]}
     \tau^{k-2}\, \Tr \left | \widehat {\mathcal J}(\tau s) \ - \
\sum_{j=1}^{k-1} \frac{1}{\tau^j} N_j(s) \right | 
 \ < \ C_k \; .
\end{equation}
This asymptotic series is derived from 
the Nenciu expansion for $\varrho_{\tau}$ \cite{nenciu}.
\item It is known that the quantization of the Hall conductance 
 is measured with very high accuracy, so that one expect that the total 
charge transport $Q(\tau)-Q(0)=\int_0^\tau\di t\mathcal J_\tau(t)$ should 
coincide with the linear response result up to higher powers of $1/\tau$ when 
$g(1)=0$. 
In the case of finite dimensional Fermi projection $P_\lambda$ this behavior 
was established rigorously in \cite{KS}. It can be also proven in our 
setup, using the approach of Avron, Seiler, and Yaffe rather then using the 
Nenciu expansion. 
\end{list}

In order to facilitate the proof of the theorem above we will use a {\it 
scaled time} $s=t/\tau$. Notice that the scaled time $s$ changes from 
$0$ to $1$ when $t$ changes from $0$ to $\tau$. Moreover we will work in the
so called interaction picture; Under the time-dependent gauge transformation
\begin{equation} \label{eq:gauge}
P_\tau(s) \ = \ \e^{\im \phi(s) \Lambda_1}  \varrho_\tau(\tau s)
\e^{-\im \phi(s) \Lambda_1} 
\end{equation}
the evolution (\ref{eq:thmheis}) is translated into the
initial value problem:
\begin{equation}\label{PtauHeis}
\begin{cases}
\im \dot P_\tau(s)  \ = \ \tau \com{H(s)}{ P_\tau(s) }
\\ \, P_\tau(0) \ = \ P_\lambda
\end{cases} \;  ,
\end{equation}
with $\phi(s) = \int_0^s g(u) \, du$  and the time dependent
Hamiltonian  $H(s) $ defined by:
\begin{equation}\label{gauge}
H(s) \ = \ \e^{\im \phi(s) \Lambda_1} \, H_\lambda \, \e^{-\im
\phi(s) \Lambda_1} \; .
\end{equation}

The utility of the working with $H(s)$ in place of
$H_\lambda(s) = H_\lambda + 1/\tau g(s) \Lambda_1$ is related to the
iso-spectrality of the former family of Hamiltonians. 

After these transformations, the current (\ref{eq:deltaJ}) becomes 
\begin{equation}\label{eq:newcurr}
\begin{split}
\Tr \widehat {\mathcal J}(\tau s) \ =& \ - \im \Tr (\varrho_\tau(\tau s) -
P_\lambda) \com{H_\lambda }{ \Lambda_2}
\\ =& \ -\im \Tr \e^{-\im \phi(s) \Lambda_1} (P_\tau(s) - P(s))
\com{H(s)}{ \Lambda_2} \e^{\im \phi(s) \Lambda_1} \\ =& \ - \im
\Tr ( P_\tau(s) - P(s)) \com{ H(s)}{ \Lambda_2} \; ,
\end{split}
\end{equation}
where
\begin{equation}\label{eq:Pdefn} P(s) \ := \ \e^{\im \phi(s)
\Lambda_1} \, P_\lambda \, \e^{-\im \phi(s) \Lambda_1} \ = \
\chi(H(s) < E_F) \; .
\end{equation}

In order to facilitate the writing of inequalities we shall adopt
the following convention: $C_{n,m,..}$ will denote a general
constant (not necessary the same at different occurrences), 
which depends only on the integers $n,m,..$, on the Sobolev norm of the 
potential $V$, and on the strength of magnetic field $B$.
\section{Asymptotic expansion for $P_\tau$ and the Kubo formula}
\label{sec:Nenciu}

In 1993 G. Nenciu \cite{nenciu} found a general form of the 
solution of the Heisenberg equation 
\begin{equation}\label{eq:motion} \im \dot P_\tau(s) \ = \ \tau
[H (s), P_\tau(s)] \, ,
\end{equation}
where $P_{\tau} (0)$ is a spectral projection of the operator $H (0)$. The idea
was to look for an asymptotic series of the form 
\begin{equation}\label{eq:asymp}
P_\tau(s) \sim B_0(s) \ + \ \frac{1}{\tau} B_1(s) \ + \
\frac{1}{\tau^2} B_2(s) \ + \ \ldots \; .
\end{equation}
The substitution of (\ref{eq:asymp}) into (\ref{eq:motion}) leads 
to a sequence of differential equations
\begin{subequations}\label{eq:nenciuproblem}
\begin{equation}\label{eq:nenciuproblema}
\im \dot B_j(s) \ = \ \com{H (s)}{B_{j+1}(s)} \quad j=0,
\ldots
\end{equation}
In addition, using that $P_\tau(s)$ is a projection for each $s$,
we get $P_\tau(s)^2 = P_\tau(s)$, which generates the following sequence 
of algebraic relations:
\begin{equation}\label{eq:nenciuproblemb}
B_j(s) \ = \ \sum_{m=0}^{j} B_m(s) B_{j-m}(s) \quad j=0, \ldots 
\end{equation}
\end{subequations}
In particular: $B_0(s)^2 = B_0(s)$, so $B_0(s)$ is a projection for
each $s$.

It turns out that the system of hierarchical relations 
\eqref{eq:nenciuproblema} and
\eqref{eq:nenciuproblemb} has a unique solution, which is given by the 
following recursive construction:
\begin{equation}\label{eq:defB}
\begin{cases}
B_0 (s) &= P(s) \\
B_{j}(s)  &=   \frac{1}{2 \pi} \int_{\Gamma} R_z(s) \com{
P(s)}{\dot B_{j-1}(s)} R_z (s) \di z + \ S_j(s) - 2 P(s) S_j(s)
P(s) \; ,
\end{cases}
\end{equation}
where $R_z(s) = (H(s) - z)^{-1}$,
\begin{equation}
S_j(s) \ = \ \sum_{m=1}^{j-1} {B_m(s) B_{j-m}(s)} \; ,
\end{equation}
and the contour $\Gamma$ encircles the spectrum below the Fermi
energy. In particular the first order (and most prominent for the linear
response) term is given by 
\begin{equation}\label{eq:B1}
B_1(s) \ =  \frac{1}{2 \pi} \int_\Gamma  R_z(s) \com {P(s)}{\dot
P(s)} R_z(s) \di z \; .
\end{equation}
One can truncate the expansion (\ref{eq:asymp}) at some finite order $k>0$ by
observing that\footnote{We owe this observation to Jeff Schenker.}:
\begin{equation}\label{eq:fin_ord}
P_\tau(s) \ = \ B_0(s)       + \frac{1}{\tau} B_1(s) \ + \
\ldots \ + \ \frac{1}{\tau^{k}} B_k(s)  - \ \frac{1}{\tau^{k}}
\int_0^s U_\tau(s,r)\dot B_k(r) U_\tau(r,s)\di r \,.
\end{equation}
where $U_\tau(s,t)$ are the Schr\"odinger  unitary propagators,
satisfying
\begin{equation}
\begin{cases}
\im \frac{\partial \phantom s}{\partial s} U_\tau(s,r) \ = \ \tau
H(s) U_\tau(s,r) \\ U_\tau(s,s) \ = \ \1 \; .
\end{cases}
\end{equation}
The observable whose trace gives the induced current can thus be expanded
according to 
\begin{equation}\label{eq:expJ}
\begin{split}
i\tau \widehat {\mathcal J}(\tau s) = &\; B_1 (s) [H(s) , \Lambda_2] + \frac{1}{\tau} B_2
(s) [H(s) , \Lambda_2] + \dots + \frac{1}{\tau^{k-1}} B_k (s) [H(s) ,
\Lambda_2] \\ &+ \frac{1}{\tau^{k-1}} \int_0^s \di r 
U_\tau(s,r)\dot B_k(r) U_\tau(r,s) [H(s) , \Lambda_2] .
\end{split}
\end{equation}
As we will see below, the first term of this expansion
exactly yields the Kubo formula 
\begin{equation}
\Tr B_1(s) \com{H(s)}{\Lambda_2} \ = \ g(s) \Tr P_{\lambda}
\com{\com{P_{\lambda}}{\Lambda_1}}{\com{P_{\lambda}}{\Lambda_2}} \;
\end{equation}
and directly proves the first part of Theorem~\ref{thm}, provided that 
we can control the other terms in the expansion (\ref{eq:expJ}), 
showing that their contribution
vanishes in the limit $\tau \to \infty$. Most of the paper deals in fact with
the control of these terms. In the following we illustrate the strategy and 
state the propositions used at the end of this section to prove Theorem
\ref{thm}. In order to keep the attention of the readers on the main physical 
ideas we defer the quite technical proofs of these propositions to later
sections.

First, we need to derive some bounds, applicable for a broad class of 
time dependent Hamiltonians with
sufficiently smooth potential. These bounds will be very useful 
to handle products of functions of the Hamiltonian with different 
functions of the space and the momentum coordinates. From here on 
we use the notations $\ax = \sqrt{1 + x^2}$ and ${\bf p_A} = ({\bf p}- {\bf A} 
( {\bf x}))$, with ${\bf A} (x_1, x_2) = B/2 (- x_2, x_1)$. 
\begin{prop}[Generalized space--momentum bounds]\label{prop:space}
Consider the Hamiltonian $H_t = p_A^2 + W(t)$ acting on $L^2 (\R^2, dx)$, 
where $W(t)$ is a time dependent multiplication operator. Fix $m,n \in \Z/2$ and
assume that $W(t) \in H_{2|m|, \infty} (\R^2)$ for all $t\in \R$ so that 
$D_m = \sup_{t \in \R} \| W(t)\|_{2|m|, \infty} <\infty$. Without loss of
generality we can also assume that $\inf_{t\in \R} \inf \sigma (H_t) \geq 1$. 
\begin{itemize}
\item[a)] There is a constant $C_m$, depending on $B$ and on $D_m$, 
such that, for any function $f \in H_{2|m|, \infty} (\R^2)$ we have
\begin{align}
\text{i) } &\| H_t^{m-1/2} {\bf p_A} H_t^{-m}\| \leq  C_m, \\
\text{ii) }  &\| H_t^{m} f(x) H_t^{-m} \| \leq  C_m \| f \|_{2|m|,
\infty}. 
\end{align}    
\item[b)] There is a constant $C_{n,m}$, depending on B and on $D_m$ such that, 
for $i=1,2$,
\begin{equation}\label{eq:lm2}
\| H_t^{m} \axi^n H_t^{-m} \axi^{-n} \| \leq C_{n,m}.
\end{equation} 
\end{itemize}
All these bounds are uniform in $t \in \R$. 
\end{prop}

\noindent {\it Remarks}
\begin{list}{(\arabic{list})}{\usecounter{list}
\setlength{\leftmargin=25pt} \setlength{\rightmargin=20pt}
\setlength{\labelsep=10pt} \setlength{\itemindent=10pt}
\setlength{\itemsep=5pt}}
\item Proposition \ref{prop:space} holds true also for $m\in \Z/4$, if we assume that
$f, W(t) \in H_{2\tilde{m}, \infty} (\R^2)$ for all $t \in \R$ and that 
$\sup_{t\in \R} \| W
(t)\|_{2\tilde{m}, \infty} < \infty$, where $\tilde{m}$ is the smallest half
integer larger or equal to $|m|$. The proof is
then very similar to the one given below, and is omitted here.
\item In particular the result applies to the Hamiltonian $H_{\lambda} (t)$ 
defined in Eq.~(\ref{eq:11}), if the potential $V$ has the required smoothness assumptions.   
\item If $H(s) = e^{i\phi(s) \Lambda_1} H_{\lambda} e^{-i\phi(s)\Lambda_1}$ is
the gauged Hamiltonian introduced in Section \ref{sec:main}, we also have  
\begin{align}
&\| H(s)^m f(x) H(s)^{-m}\| \leq C_m \| f \|_{2|m|, \infty} \quad \text{and} \\
&\| H(s)^m \axi^n H(s)^{-m} \axi^{-n} \| \leq C_{n,m}\,,
\end{align}
because $e^{i\phi(s)\Lambda_1}$ commutes with the operators
$f(x)$ and $\axi$. 
\end{list}

According to Eq. (\ref{eq:expJ}) we have to control 
two different types of terms
\begin{equation}\label{eq:trace1}
\frac{1}{\tau^{j-1}}\Tr B_j(s) \com{H(s)}{\Lambda_2} \quad \text{with } 2\le j\le k,
\end{equation} 
and 
\begin{equation}\label{eq:trace2}
\frac{1}{\tau^{k-1}}\Tr \int_0^s \di r \, U_\tau(s,r)\dot B_k(r) U_\tau(r,s)
\com{H(s)}{\Lambda_2}\,. 
\end{equation}
In order to estimate these traces we use the following result.

\begin{prop}[Trace class estimates]\label{prop:tr}
Suppose that the operators $A,B$ acting on $L^2 (\R^2, dx)$ satisfy the two
conditions 
\begin{equation}\label{trnor}
\|A\, \axvec^{3}\| <\infty\,;\quad \|H^{3/2}_\lambda\, D\| <\infty\,.
\end{equation}
Then $AD$ is a trace class operator.
\end{prop}

Using the last proposition the problem of showing that the traces
(\ref{eq:trace1}) and (\ref{eq:trace2}) are finite reduce to the problem of
proving that the corresponding operators decay sufficiently fast in the energy
and in the space coordinates. Consider first the term that appears in
Eq.~(\ref{eq:trace1}). The corresponding decay property follows from the fact that the 
operator $B_j (s)$ is localized in the energy and in the $x_1$ coordinate, 
while the commutator $[H(s),\Lambda_2]$ decays in the $x_2$ coordinate. Here are the corresponding claims. 

\begin{prop}[Fast decay away from the $x_2$ axis]\label{prop:bj}
Fix $N \in \N$ and assume that $V \in H_{2N, \infty} (\R^2)$. Then, for all
$j=1,2,\dots $ we have  
\begin{equation}
\sup_{s\in [0,1]} \|B_j(s)\, H^N(s) \axone^{N}\| < C_{j,N}\;,\label{trnor1}
\end{equation} and for all $j=0,1,\dots $
\begin{equation}
\sup_{s\in [0,1]} \|\dot B_j(s) \, H^N(s) \axone^{N}\| < C_{j,N}\,, \label{trnor2}
\end{equation}
\end{prop}

\begin{prop}[Localization of the current operator near the $x_1$ axis]\label{prop:cur}
The operator $\com{H(s)}{ \Lambda_2}$ is supported in a strip of
width $2m$ around $x$ axis. Moreover, if $V \in H_{1, \infty} (\R^2)$, we have 
$$\|H^{-1/2}(s)\,\axtwo^{N}\,\com{H(s)}{ \Lambda_2}\|\ \le \ C_N\,.$$
\end{prop}
\noindent {\it Remark} The proof of this claim is a trivial consequence of the
locality of the Hamiltonian $H(s)$ and Proposition \ref{prop:space}.
 
Propositions \ref{prop:tr}-\ref{prop:bj} imply that the trace 
\[ \Tr B_j(s) [H(s), \Lambda_2] \] is finite and since it is $\tau$-independent, 
we observe that each term (\ref{eq:trace1})
vanishes in the limit $\tau \to \infty$. What about the remainder 
(\ref{eq:trace2})? Also in this case the operator $\dot{B}_k
(r)$ decays in the energy and in the $x_1$ coordinate, and the commutator
$[H(s), \Lambda_2]$ decays in the $x_2$ coordinate. But the two operators
are separated by the time evolution $U_{\tau} (r,s)$. Knowing that 
$[H(s), \Lambda_2]$ is localized in a strip of length $2m$ around 
the $x_1$ axis, what can be said
about $U_{\tau} (r,s) [H(s), \Lambda_2]$? To answer this question we
use the fact that the electrons cannot propagate faster than ballistically. 
Since the energy of the electrons is essentially bounded by the Fermi energy (
modulo the spread due to time dependent potential), we observe that electrons 
initially confined inside the strip of width $2m$ can, in the course of their 
evolution, propagate onto a strip of width $O(\tau)$ (because 
the time difference $|r-s|$ can be of order $\tau$) around 
the $x_1$ axis. This is the content of 
Proposition \ref{prop:speed} below. Using this result we can then 
prove that the operator 
\begin{equation}
\int_0^s \di r \, U_\tau(s,r)\dot B_k(r) U_\tau(r,s)
\com{H(s)}{\Lambda_2}\,
\end{equation}
is trace class, and that the corresponding norm is 
proportional to some power of $\tau$. Thus choosing the order $k$
of the expansion (\ref{eq:expJ}) sufficiently large, also the term
(\ref{eq:trace2}) vanishes in the limit $\tau \to \infty$. 

\begin{prop}\label{prop:speed}[Finite speed of propagation]
Consider the Hamiltonian $H_{\lambda} (t)$ as defined in Eq. (\ref{eq:11}) 
and denote by $U(t,s)$ the
corresponding Schr\"odinger evolution. 
Fix $n,m \in \N/2$, and assume that $V \in H_{2(n+m),\infty} (\R^2)$. 
Then there is a constant $D=D(n,m)$ such that, for $i=1,2$,  
\begin{equation}
\| \axi^n H_{\lambda} (t_1)^m U(t_1,t_2) H_{\lambda} 
(t_2)^{-m-n} \ph \| \leq D \langle t_1-t_2 \rangle^{n}  
\| \axi^n \ph \| ,
\end{equation}
for all $t_1 ,t_2 \in \R$. 
\end{prop}
\noindent {\it Remarks}
\begin{list}{(\arabic{list})}{\usecounter{list}
\setlength{\leftmargin=25pt} \setlength{\rightmargin=20pt}
\setlength{\labelsep=10pt} \setlength{\itemindent=10pt}
\setlength{\itemsep=5pt}}
\item The proposition actually holds also for $m \in \Z/2$. 
The proof in this case is identical to the one given below 
for the case $m \geq 0$ and is therefore omitted. 
\item It follows from this proposition that, for all $n\in \N/2$ and $m \in
\Z/2$, \[ \| \axi^n H_{\lambda} (t_1)^m U(t_1,t_2) H_{\lambda} (t_2)^{-m-n} 
\axi^{-n} \| \leq D \langle t_1-t_2 \rangle^{n} .\]
\item After rescaling the time $s =t/ \tau$ and introducing 
the gauged Hamiltonian $H(s) = e^{i \phi (s)
\Lambda_1} H_{\lambda} e^{-i\phi (s) \Lambda_1}$ the last equation 
implies that, for $0 < s_1 , s_2 <1$, and for all $n \in \N/2$ and 
$m \in \Z/2$, we have  
\begin{equation}
\| \axi^n H (s_1)^m U_{\tau} (s_1 , s_2) H (s_2)^{-m-n} \axi^{-n} \| 
\leq D \langle \tau \rangle^{n},
\end{equation}
where $U_{\tau} (t,s)$ denotes the time evolution generated by $H(s)$. 
This is the bound explicitly used in the applications. 
\end{list}

Using Propositions 1-5 and following the strategy outlined above, we can now
prove our main result, Theorem \ref{thm}.

\begin{proof}[Proof of Theorem \ref{thm}] Part
(1): We use the expansion (\ref{eq:expJ}) of the operator $\tau \widehat 
{\mathcal J} (s\tau)$, and we estimate the trace of the different terms in the expansion.
Clearly, 
\begin{equation}
\begin{split}
\|B_j\com{H(s)}{ \Lambda_2}\|_1 \leq \; &\|B_j(s)\, H^N(s) \bx^{N}\| 
\|\bx^{-N} H^{-N}(s)\cx^{-N} H^{1/2}(s)\|_1 \\ &\cdot 
\|H^{-1/2}(s)\cx^{N}\com{H(s)}{
\Lambda_2}\|\,.
\end{split}
\end{equation}
The first and the last factors are finite by Propositions
\ref{prop:bj} and \ref{prop:cur}. Since
\begin{equation}
H(s) \ = \ \e^{\im \phi(s) \Lambda_1} \, H_\lambda \, \e^{-\im
\phi(s) \Lambda_1} \; ,
\end{equation}
we obtain that
\begin{equation}\label{eq:s-l}
\|\bx^{-N}H^{-N}(s) \cx^{-N}H^{1/2}(s)\|_1  = 
\|\bx^{-N}H^{-N}_\lambda \cx^{-N}H^{1/2}_\lambda\|_1\,.
\end{equation}
Proposition \ref{prop:space} implies in turn that
\begin{equation}\label{eq:opo}
\begin{split}
\|\bx^{-N}&H^{-N}_\lambda \cx^{-N} H^{1/2}_\lambda\|_1  \\ \leq \; & 
\|\bx^{-N}H^{-N+1/2}_\lambda \cx^{-N}\|_1 \|\cx^{N}H^{-1/2}_\lambda
\cx^{-N}H^{1/2}_\lambda\|\\ \leq \; &C_N\,\|\bx^{-N}H^{-N+1/2}_\lambda
\cx^{-N}\|_1\\ \leq \; & C_N\,
\|\bx^{-N}\cx^{-N}H^{-N+1/2}_\lambda\|_1 
\|H^{N-1/2}_\lambda \cx^{N}H^{-N+1/2}_\lambda \cx^{-N}\| \\ \leq \; &
C_N  \, \|\axvec^{-N}H^{-N+1/2}_\lambda\|_1\,.
\end{split}
\end{equation}
The latter norm is finite by Proposition \ref{prop:tr}.

Now we need to estimate the trace of the last term in the expansion
(\ref{eq:expJ}), 
\begin{equation}\label{eq:rem2}
\frac{1}{\tau^{k-1}} \int_0^s \di r 
U_\tau(s,r)\dot B_k(r) U_\tau(r,s) [H(s) , \Lambda_2] .
\end{equation}
To this end we note that 
\begin{equation*}
\begin{split}
\|U_\tau(s,r)&\dot B_k(r) U_\tau(r,s)\com{H(s)}{ \Lambda_2}\|_1 \\ \leq \; &
\|\dot B_k(r) \, H^N(r) \bx^{N}\| \, \|\bx^{-N}H^{-N}(r)\cx^{-m}H^{m + 1/2}(r)\|_1 \\ 
\cdot \; &\|H^{-m - 1/2}(r)\cx^{m} U_\tau(r,s) \cx^{-m} H^{1/2}(s)\| \,
\|H^{-1/2}(s)\cx^{m}\com{H(s)}{ \Lambda_2}\|\,.
\end{split}
\end{equation*}
The first and the forth factors are bounded by
Propositions \ref{prop:bj} and \ref{prop:cur}, accordingly. The third term is
bounded by $C_m \tau^{m}$, because of Proposition \ref{prop:speed}. Choosing
$N=2m$, and $m \geq 3$ the second term is bounded by Propositions \ref{prop:tr} and 
\ref{prop:space} (this can be shown as in Eq.(\ref{eq:opo})).This implies that
the absolute value of the trace of (\ref{eq:rem2}) is bounded by 
$C_{k,m} \tau^{m-(k-1)}$, for any $m \geq 3$ and for any $k \geq 1$. 
Thus choosing $k \geq 5$ and $m=3$ it follows that the trace of $\tau
\widehat{\mathcal{J}} (s\tau)$ is finite, uniformly in $\tau$ and that 
\begin{equation}
| \tau \mathcal{J} (s\tau) + i \Tr B_1 (s) [H(s) , \Lambda_2] | \leq C
\tau^{-1} 
\end{equation}
Part (2): We need to prove that 
\begin{equation}
\label{eq:suffforKubo} \Tr B_1(s) \com{H(s)}{\Lambda_2} \ = \ g(s)
\Tr P_{\lambda} \com{\com{P_{\lambda}}{\Lambda_1}}{\com{P_{\lambda}}{\Lambda_2}} \; .
\end{equation}
This relation follows from the explicit equation for $B_1(s)$ and the 
cyclicity of 
the trace. Indeed,
\begin{equation}
B_1(s) \ = \ \frac{1}{2 \pi} \int_\Gamma  R(s,z) \com
{P(s)}{\com{P(s)}{\im
g(s) \Lambda_1}} R(s,z) \di z \; .
\end{equation}
One may now verify Eq.~(\ref{eq:suffforKubo}) using that 
\begin{equation}
\begin{aligned}
\Tr B_1(s) \com{H(s)}{\Lambda_2} \ = & \ \frac{\im g(s) }{2
\pi}\int_\Gamma \di z
   \, \Tr
\com{P(s)}{\com{P(s)}{\Lambda_1}}  R_z (s) \com{ H(s)}{\Lambda_2}
R_z (s) \\ =& \ \frac{\im g(s) }{2 \pi}\int_\Gamma \di z \, \Tr
\com{P(s)}{\com{P(s)}{
\Lambda_1}} \com{\Lambda_2}{R_z(s)}  \\
=& \ g(s) \, \Tr
\com{\com{P(s)}{ \Lambda_1}} {P(s)} \com{\Lambda_2}{P(s)} \; ,
\end{aligned}
\end{equation}
since $\com{P(s)}{\com{P(s)}{\Lambda_1}}  R_z(s)
\com{H(s)}{\Lambda_2}$ is trace class, as follows from the proof of the first 
part of Theorem 
\ref{thm} (note that $g(s)\com{P(s)}{\Lambda_1}=\dot B_0(s)$). 
It is also immediate from the proof of the first part that the product 
$\com{P(s)}{\Lambda_1}\com{\Lambda_2}{P(s)}$ is also trace 
class. Therefore, the cyclicity of
the trace  yields
\begin{equation}
\Tr B_1(s) \com{H(s)}{ \Lambda_2} \ =
   \ g(s) \Tr P(s) \com{\com{P(s)}{
\Lambda_1}}{\com{P(s)}{\Lambda_2}} \; ,
\end{equation}
which is equivalent to \eqref{eq:suffforKubo} since
$\e^{\pm \im \phi(s) \Lambda_1}$ commutes with $\Lambda_1$ and
$\Lambda_2$; recall that $P(s) = \e^{\im \phi(s) \Lambda_1} P_{\lambda}
\e^{-\im \phi(s) \Lambda_1}$. 

Part (3) - the stability problem. 
We are going to employ here the following tactics: First, we will
demonstrate that $K_\lambda$ is stable with respect to changes in the potential
far away from the origin. Secondly, we show that the
relative trace class perturbation - namely the change of the potential in
a finite region around the origin -  also leaves $K_\lambda$ invariant. The first
part will follow from the first resolvent identity:
\begin{equation}
(H_\lambda-z)^{-1}-(\hat H_\lambda-z)^{-1} \ = \
(H_\lambda-z)^{-1}\lambda V\chi_L(\hat H_\lambda-z)^{-1}\,,
\end{equation}
where $1-\chi_L$ is a smooth characteristic function of the ball
of radius $L$, centered at the origin, and $\hat
H_\lambda:=H_\lambda -\lambda V\chi_L$. Let us compare now
$K_\lambda$ and $\hat K_\lambda$, where the latter is the expression of the 
Kubo formula computed for $\hat H_\lambda$:
\begin{multline}\label{fb}
K_\lambda-\hat K_\lambda=\Tr (P-\hat P)
\com{\com{P}{\Lambda_1}}{\com{P}{\Lambda_2}}\\ + \Tr \hat P
\com{\com{(P-\hat P)}{\Lambda_1}}{\com{P}{\Lambda_2}} \\ + \Tr
\hat P \com{\com{\hat P}{\Lambda_1}}{\com{(P-\hat P)}{\Lambda_2}}\,.
\end{multline}
Here we use the concise notation $P$ instead of $P_{\lambda}$ and $\hat{P} =
\chi (\hat{H}_{\lambda} \leq E_F)$. The idea is that $\com{P}{\Lambda_1}\com{P}{\Lambda_2}$ is
basically supported near the origin, while $P-\hat P$ is
essentially supported outside the ball of radius $\sqrt{L}$, hence their
product is small. This idea can be materialized in the following fashion: Since 
$$P-\hat P=\int_\Gamma \di z \hat R_z\lambda V\chi_LR_z$$ and 
\begin{multline}
\|(1-\chi_{\sqrt L})R_z\chi_L\|\\ \le\ \|(1-\chi_{\sqrt
L})<x>^{2N}\|\,\|<x>^{-2N}R_z<x>^{2N}\|\,\|<x>^{-2N}\chi_L\|\\
\le \ \frac{1}{\dist(z,\sigma (H_\lambda))}C_{N}L^{-N}\,,\nonumber
\end{multline}
we get the bound 
$$\|(P-\hat P)
\com{\com{P}{\Lambda_1}}{\com{P}{\Lambda_2}}\|_1\ \le \  
\|\chi_{\sqrt L}\com{P}{\Lambda_1}\com{P}{\Lambda_2}\|_1 \ + \ 
C_{N}L^{-N}\,.$$
On the other hand,
\begin{multline}
\|\chi_{\sqrt L}\com{P}{\Lambda_1}\com{P}{\Lambda_2}\|_1\\ \le \|
\chi_{\sqrt L}<x>^{-2N}\|\, \|
<x>^{2N}\com{P}{\Lambda_1}\com{P}{\Lambda_2}\|_1 \ \le  \ C_NL^{-N}\,,
\end{multline}
where we used Proposition \ref{prop:bj} (note that $\dot B_0(s)= 
\dot P(s)= g(s)\com{P(s)}{\Lambda_1}$).

Hence the trace norm of the first contribution in Eq. (\ref{fb}) is
bounded by $ C_N L^{-N}$. Literally the same bounds holds whenever
one replace $P$ by $\hat P$, therefore all contribution on the r.h.s.
of Eq. (\ref{fb}) are bounded by the same bound, and choosing $L$
large enough, one can make difference $K_\lambda-\hat K_\lambda$
arbitrarily small.

Next, we want to show that $\hat K_\lambda$ is independent of
$\lambda$. For this purpose let us compute its derivative with
respect to $\lambda$. It is convenient to rewrite $\hat
K_\lambda$ as 
$$\hat K_\lambda = \Tr
\com{\hat P\Lambda_1\hat P}{\hat P\Lambda_2\hat P}\,.$$ Since 
the expression under the trace is a commutator, so that the
derivative of $\hat K_\lambda$ is zero if (a) the operator inside
the trace is trace class for any coupling in the vicinity of
$\lambda$ and we can interchange the trace and the
derivative; (b) if $\partial_\lambda \hat P$ is trace class. The
first item is a consequence of the bounds
$$\|\com{\hat{P}}{\Lambda_1}\, H^N(s) <x_1>^{N}\|\le C_N\,;\quad \|\com{\hat{P}}
{\Lambda_2}\, H^N(s) <x_2>^{N}\|\le C_N\,,$$ which follow from Proposition 
\ref{prop:bj} and from the complete symmetry between $x_1$ and $x_2$. 
To prove the second
one, observe that 
$$\partial_\lambda \hat P=-\int_\Gamma \di z \hat R_z V(1-\chi_L) \hat
R_z\,.$$ Since $\hat R_z \hat H$ is uniformly bounded  for all
$z\in\Gamma$, the trace class condition can be seen from the
boundness of $HV(1-\chi_L)H^{-1}$ and Proposition \ref{prop:tr}.
Since the difference between $\hat K_\lambda$ and
$K_\lambda$ can be made arbitrarily small, we conclude the result.
\end{proof}

\section{Phase space bounds and Trace Estimates}\label{sec:tr}

In this section we will prove Propositions \ref{prop:space}-\ref{prop:bj}. 

\begin{proof}[Proof of Proposition \ref{prop:space}]
a) It is enough to check the bounds for $m \geq 0$. The proof is then 
by induction over $m$. For $m=0$ both bounds i) and ii) are obvious. 
We assume now that i) and ii) hold true for all $m 
\leq M-1/2$, and we prove the statements for $m=M$. 
If $M=1/2$, i) is clear. For $M\geq 1$ we have, using the concise 
notation $H\equiv H_t$,  
\begin{equation}
 H^{M-1/2} {\bf p_A} H^{-M} = H^{M-3/2} {\bf p_A} H^{-M+1} + H^{M-3/2} [H, {\bf 
 p_A}] H^{-M}. 
\end{equation}
Here \([{\bf p_A},H] = iB {\bf \tilde{p}_A} + i {\bf \nabla} W (t) \), 
where ${\bf \tilde{p}_A} = (p_{A,y} , -p_{A,x})$ if ${\bf p_A} = (p_{A,x} , p_{A,y})$. 
It follows that 
\begin{equation}
\begin{split}
H^{M-1/2} {\bf p_A} H^{-M} = \; &H^{M-3/2} {\bf p_A} H^{-M+1} + iB H^{M-3/2}{\bf
\tilde{p}_A } H^{-M} \\ &+ i H^{M-3/2} {\bf \nabla} W (t) H^{-M} 
\end{split}
\end{equation}
and thus that
\begin{equation*}
 \| H^{M-1/2} {\bf p_A} H^{-M} \| \leq (1+B)\, \|H^{M-3/2} {\bf p_A} H^{-M+1}\| + 
 \| H^{M-1} {\bf \nabla} W (t) H^{-M+1} \|. 
\end{equation*}
Applying the induction hypothesis with $m=M-1$ (since $W(t) 
\in H_{2m, \infty} (\R^2)$ we have ${\bf \nabla} W (t) \in H_{2m-1, \infty} 
(\R^2) \subset  H_{2m-2,\infty} (\R^2)$) it follows that 
\begin{equation*}
\|H^{M-1/2} {\bf p_A} H^{-M}\| \leq C_M
\end{equation*}
for a constant $C_M$ depending only on $B$ and on 
$D_M = \sup_{t \in \R} \| W (t) \|_{2M,\infty}$. This proves
part i). As for part ii) for $m=M$,  we consider first the case $M=1/2$.
Then we have 
\begin{equation}
H^{1/2} f(x) H^{-1/2} = f(x) + \frac{1}{\pi} \int_0^{\infty} \frac{ds}{\sqrt{s}}
\frac{H^{1/2}}{H+s} [f(x), H] \frac{1}{H+s}. 
\end{equation} 
Since $[f(x),H] = 1/2 ({\bf p_A} \cdot {\bf \nabla} f + {\bf \nabla} f \cdot {\bf p_A} )$ we find 
\begin{equation}
\begin{split}
\|H^{1/2} f(x) H^{-1/2}\| \leq \; &\| f(x)\| + 
\frac{1}{2\pi} \int_0^{\infty} \frac{ds}{\sqrt{s}}
\| \frac{H}{H+s}\| \| H^{-1/2} {\bf p_A} \| \| \nabla f \| \| \frac{1}{H+s}\| \\ 
&+\frac{1}{2\pi} \int_0^{\infty} \frac{ds}{\sqrt{s}}
\| \frac{H^{1/2}}{H+s}\| \| \nabla f \| \| {\bf p_A} H^{-1/2}\| \|
\frac{H^{1/2}}{H+s}\| \\
\leq \; &C \| f \|_{1, \infty} 
\end{split}
\end{equation} 
where the constant $C$ depends only on $D_0 = \sup_{t \in \R} 
\| W (t)\|_{0, \infty}$. This proves ii) if $M=1/2$. Now we assume $M\geq 1$. 
In this case we have
\begin{equation}\label{eq:pe1,3}
\begin{split}
H^{M} f(x) H^{-M} &= H^{M-1} f(x) H^{-M+1} + H^{M-1} [H, f(x)] H^{-M} \\
&= H^{M-1} f(x) H^{-M+1} + \frac{1}{2} H^{M-1} \left({\bf p_A} \cdot {\bf \nabla
}f + \nabla
f \cdot {\bf p_A} \right) H^{-M}, 
\end{split}
\end{equation}
which implies that
\begin{equation}
\begin{split}
\| H^{M} f(x) H^{-M} \| \leq \; &\|H^{M-1} f(x) H^{-M+1} \| +  
\| H^{M-1/2} \nabla f H^{-M+1/2} \| \\ &\times \frac{1}{2} \left( \| H^{M-1} 
{\bf p_A} H^{-M+1/2} \| + \|H^{M-1/2} {\bf p_A} H^{-M} \|\right).  
\end{split}
\end{equation}
Using the induction hypothesis with $m=M-1/2$ and $m=M-1$, and using i) 
for $m=M$ (which was proven above) we find
\begin{equation}
\| H^{M} f(x) H^{-M} \| \leq C_M \|f \|_{2M, \infty},
\end{equation}
where $C_M$ depends on $B$ and on $D_M = \sup_{t \in \R} \| W \|_{2M, \infty}$. 
Here we used that $\| \nabla f \|_{2M-1, \infty} \leq \|f \|_{2M, \infty}$. 
This completes the proof of the claim a).

b) We assume that $m,n \geq 0$,  the other values can be treated similarly. We
consider first the case $m\in \N$ (while $n$ can also be half-integer) and we
prove the claim by induction over $m$. For $m=0$ the result
is trivial. Now we assume it holds true for $n\in \N/2$ and $m\leq M-1$, 
and we prove it for $m=M$ (and all $n\in \N/2$). 
To this end we proceed by induction over $n$. For $n=0$ the claim is again trivial. Thus we
assume it holds also if $m=M$ and $n\leq N-1/2$, for some $N\in \N/2$.
Then we have  
\begin{equation}\label{eq:lm2-ind}
\begin{split}
H^M &\axi^N H^{-M} \axi^{-N} = \; H^{M-1} \axi^N H^{-M+1} \axi^{-N} \\ &+ 
H^{M-1} [H, \axi^N] H^{-M} \axi^{-N} 
=  H^{M-1} \axi^N H^{-M+1} \axi^{-N} \\ &+ 2 i  H^{M-1}  
{\bf p_A} \cdot \frac{x_i}{\axi} \axi^{N-1} H^{-M} \axi^{-N} \\ 
&+  H^{M-1} (2- \axi^{-2})\axi^{N-2}H^{-M} \axi^{-N}.
\end{split}
\end{equation}
The first term on the r.h.s. of the last equation is bounded by induction
assumption (with $m=M-1$ and $n=N$). The second 
term on the r.h.s. of (\ref{eq:lm2-ind}) is also bounded. Indeed, for $N=1/2$
this follows by 
part a) of the proposition. Otherwise we can write this term as   
\begin{equation} 
\left( H^{M-1} p_{A,i} H^{-M}\right)\left(H^{M} \frac{x_i}{\axi} H^{-M}\right)\left(
H^{M} \axi^{N-1} H^{-M} \axi^{-N}\right) 
\end{equation}
which is bounded, because $H^{M-1} p_{A,i} H^{-M}$ and $H^{M} x_i / \axi H^{-M}$ are
bounded by part a) of the proposition and because $H^{M} \axi^{N-1} H^{-M}
\ax^{-N}$ is bounded by the induction assumption.  The
boundedness of the third term on the r.h.s. of (\ref{eq:lm2-ind}) can be 
verified analogously. 
This proves part b) for all $m\in\N$, and $n\in \N/2$. To establish part b) 
for half integer values of $m$ we can apply the same 
induction argument given by Eq. (\ref{eq:lm2-ind}) and thus it only remains to
prove the result for $m=1/2$ and all $n\in \N/2$. We observe that
\begin{equation*}
\begin{split}
&H^{1/2} \axi^n H^{-1/2} \axi^{-n} = 1 + \frac{1}{\pi} \int_0^{\infty}
\; \frac{ds}{\sqrt{s}} \frac{H^{1/2}}{H+s} [ \axi^n , H] \frac{1}{H+s} \axi^{-n} \\
&= 1 + \frac{1}{\pi} \int_0^{\infty} 
\frac{ds}{\sqrt{s}} \frac{H^{1/2}}{H+s} \left( p_{A,i} \, 
\frac{x_i}{\axi} \axi^{n-1}
+ \axi^{n-1} \frac{x_i}{\axi}\, p_{A,i} \right)  
\frac{1}{H+s} \axi^{-n} .
\end{split}
\end{equation*}
Consequently,
\begin{equation}
\begin{split}
\|H^{1/2} &\axi^n H^{-1/2} \axi^{-n} \| \leq \; C_1 + C_2 \int_0^{\infty} 
\frac{ds}{\sqrt{s}} \| \frac{H^{1/2}}{\sqrt{H+s}}\| \| (H+s)^{-1/2} p_{A,i} \|
\\ &\times \|\frac{1}{H+s} \| \| (H+s) \axi^{n-1} \frac{1}{H+s} \axi^{-n}\| \\ 
&+ C_3 \int_0^{\infty} \frac{ds}{\sqrt{s}} \| \frac{H^{1/2}}{H+s}\| 
\|\frac{1}{H+s} \| \| (H+s) \axi^{n-2} \frac{1}{H+s} \axi^{-n}\| .
\end{split}
\end{equation}
Applying the proposition for $m=1$ (which, as we already checked, holds true) 
and the fact that $\| (H+s)^{-1} \| \leq (s+1)^{-1}$, last equation proves 
b) for $m=1/2$ and for all $n\in \N/2$. 
\end{proof}

Next we prove Proposition \ref{prop:tr}, which gives us a simple condition for
 an operator to be trace class. 

\begin{proof}[Proof of Proposition \ref{prop:tr}]
This result is based on two observations:
the product of two Hilbert Schmidt operators is trace class
(by the generalized H\"older inequality for trace ideals), and that the operator
$$\frac{1}{\sqrt{p^2+1}}\ax^{-1}$$ is in $\Sch_3$ in $2D$ by the Birman-Solomyak
Theorem \cite{birman}. 
The diamagnetic inequality and Proposition \ref{prop:space} imply that 
also $$\frac{1}{\sqrt{p_A^2 + 1}} \ax^{-1}\in\Sch_3\, ,$$ where
${\bf p_A} = ({\bf p} - {\bf A}({\bf x}))$. Since $H_\lambda +B \ge p_A^2$,  also
$$\frac{1}{\sqrt{H_\lambda +B + 1}}\ax^{-1}\in\Sch_3\,$$  in two
dimensions, hence
$$\left (\frac{1}{\sqrt{H_\lambda + B +1}}\ax^{-1}\right )^3$$ is trace
class. Using Proposition \ref{prop:space}, one obtain that
$$\frac{1}{(H_\lambda + B +1)^{3/2}}\frac{1}{\ax^{3}}$$ is trace class, hence
$$AD \ = \ A\,\ax^{3}\left
\{\frac{1}{\ax^{3}}\frac{1}{(H_\lambda + B +1)^{3/2}}\right \}
(H_\lambda + B +1)^{3/2} \, D$$ is trace class, using the bounds in Eq. (\ref{trnor}) 
and the fact that the trace class is a two-side
ideal upon multiplication by bounded operators.
\end{proof}

In order to verify Proposition \ref{prop:bj} we have to show
that the operators $B_j(s)$, introduced in Section \ref{sec:Nenciu}, decay in
the energy and in the $x_1$ coordinate. Looking for example at the definition
(\ref{eq:B1}) of $B_1 (s)$, we see that it contains the projection $P(s)=
\chi (H(s) \leq E_F)$, which gives the necessary decay in the energy, and also the time 
derivative $\dot{P} (s) = i g(s) [P(s) , \Lambda_1]$ which also gives the 
decay in the $x_1$ coordinate (because of the commutator $[P(s), \Lambda_1]$). 
In the proof below we explain how to make this argument precise and how to
generalize it, by induction, to the operators $B_j (s)$, for $j\geq 2$. 

\begin{proof}[Proof of Proposition \ref{prop:bj}]
We proceed by induction over $j \in \N$. We first establish 
the inequality (\ref{trnor2}) for $j=0$, namely that
\begin{equation}\label{norP}
\sup_s\|\dot P(s)\, H^N(s) \bx^{N}\| < C_{N}\,.
\end{equation}
In order to check the last inequality we use the simple identity $\dot{P} (s) H^N (s) =
\partial_s (P(s) H^N (s)) - P(s) \partial_s (H^N (s))$. Since \[ 
\partial_s\left (H^N(s)\right ) =  g(s)\sum_{l=0}^{N-1}
H^l(s)[H(s),\Lambda_1]H^{N-1-l}(s) ,\] we find 
\begin{equation}\label{eq:proof3,2} 
\begin{split}
\|\dot{P} (s) H^N (s) \bx^N \| \leq \; &\| \partial_s (P(s) H^N (s)) \bx^N \|
\\ &+
| g(s)| \sum_{k=0}^{N-1} \| P(s) H^k(s) [H(s),\Lambda_1] H^{N-1-k}(s)\bx^N
\|.
\end{split}
\end{equation}
Consider first the terms in the sum over $k$. Since $\partial_1 \Lambda_1 = 0$
if $|x_1| > m$, for some $m>0$, and because of the locality of $H(s)$, we have \(
[H(s),\Lambda_1] H^{N-1-l}(s) = [H(s),\Lambda_1] H^{N-1-l}(s) \chi (|x_1| \leq
m) \). Since $\| \chi (|x_1| \leq m) \bx^N \|$ and $\|P(s) H^N (s) \|$ are
bounded for all $N \in \N$ and uniformly in $s$, we find 
\begin{equation*} 
\| P(s) H^k(s) [H(s),\Lambda_1] H^{N-1-k}(s)\bx^N \| \leq C_N \, 
\| H^{-N + k} [H(s),\Lambda_1] H^{N-1-k}(s) \| ,
\end{equation*}
where the r.h.s. is  finite by Proposition \ref{prop:space}, part a). 
It remains to consider the first term on
the r.h.s. of (\ref{eq:proof3,2}). Here we use the integral 
representation \[ P(s) H^N(s)= \int_\Gamma \di z z^N R_z(s) \] and we find,
since $\dot{R}_z (s) = -i g (s) R_z (s) [H(s) , \Lambda_1 ] R_z (s)$, 
that    
\begin{equation*}
\begin{split}
\| \partial_s (P(s) H^N (s)) \bx^N \| \leq |g (s)| \int_{\Gamma} |dz|
|z|^N \|R_z (s) [H(s) , &\Lambda_1] \bx^N \| \\  
&\times\| \bx^{-N} R_z (s) \bx^N \|.
\end{split}
\end{equation*}
The r.h.s. of the last equation is bounded because \( [H(s) , \Lambda_1] =[H(s)
, \Lambda_1] \chi (|x_1| \leq m) \) and because, by Proposition
\ref{prop:space}, \( \|\bx^{-N} R_z (s) \bx^N \| < C_N \), uniformly in $s$
and in $z \in \Gamma$. This establishes Eq. (\ref{norP}). In order to prove 
Eq. (\ref{trnor1}) for $j=1$ we use that, by virtue of (\ref{eq:B1}), 
\begin{equation}
\begin{split}
\|B_1 (s) H^N \bx^N \| &\leq \frac{1}{2\pi} \int_{\Gamma} |dz| \|R_z (s) [P(s),
\dot{P} (s)] R_z (s) H^N (s) \bx^N \| 
\\ &\leq C \int_{\Gamma} |dz| \| \dot{P} (s) H^N (s) \bx^N \| \\& \times
\Big (\| \bx^{-N} 
P(s) R_z (s) \bx^N \| + \| \bx^{-N}  R_z (s) \bx^N \|\Big )\,,
\end{split}
\end{equation}
and that the expression in the brackets is bounded by $C_N$. Indeed, 
it follows from Proposition \ref{prop:space}, using an
integral representation of $P(s)$ in terms of $R_z (s)$. In order to verify 
Eq. (\ref{trnor2}) for $j=1$ we use that, by Proposition \ref{prop:space},  
\begin{equation}\label{eq:lambda1}
\| \bx^{-N} H^{-N} (s) \Lambda_1 H^{N} (s) \bx^N \| \leq C_N, 
\end{equation}
and that $\dot{B}_1 (s) = g (s) [\Lambda_1 , B_1 (s) ] + \dot g(s)
(g(s))^{-1}B_1 (s)$.

For general $j >1$ we have 
\begin{equation*}
B_{j}(s) \ = \ \frac{1}{2 \pi}
\int_{\Gamma} R_z(s) \com{ P(s)}{\dot B_{j-1}(s)} R_z(s) \di z 
+ \ S_j(s) - 2 P(s) S_j(s) P(s) \; ,
\end{equation*}
and
\begin{equation}
S_j(s) \ = \ \sum_{m=1}^{j-1} {B_m(s) B_{j-m}(s)} \; .
\end{equation}
Thus Eq. (\ref{trnor1}) follows directly by the induction Hypothesis and 
Proposition \ref{prop:space}. Similarly, Eq. (\ref{trnor2}) can be proven 
using the induction hypothesis because $\dot
B_{j}(s)=g(s)[\Lambda_1,B_{j}(s)]+\tilde{B}_{j}(s)$, where $\tilde{B}_j
(s)$ has the same structure as $B_j (s)$.
\end{proof}

\section{Propagation Estimate}\label{sec:propag}

In order to prove Proposition \ref{prop:speed} we first need an auxiliary
result, which ensures that the energy remains bounded during the physical evolution. We
first learned about the existence of such bounds from Gian Michele Graf. 

\begin{lemma}[Energy boundedness]\label{lm:energy}
Suppose $H_{\lambda} (t)$ is as in Eq.~(\ref{eq:11}) and let
$U(t,s)$ be the time evolution generated by $H_{\lambda} (t)$. Then 
\begin{equation}
\sup_{s,t \in \R} \| H_{\lambda}^{-m/2} (s) U(s,t) H_{\lambda}^{m/2}(t)\|
\le C_m
\end{equation}
for all integer values of $m$.
\end{lemma}

\begin{proof}
We use the gauged transformed Hamiltonians $H(s)=e^{i\phi(s)\Lambda_1}
H_{\lambda} e^{-i\phi (s) \Lambda_1}$ (where $s$ is the scaled time) 
and the time evolution $U_{\tau} (s,t)=
e^{i\phi(s)\Lambda_1} U(s,t) e^{-i\phi (t) \Lambda_1}$ generated by the 
Hamiltonians $H(s)$. Moreover since the $H(s)$ are constant for $s >1$, it is
enough to prove that 
\begin{equation} 
\sup_{s,t \in[0,1]} \|H^{-m/2}(s)U_\tau(s,t)H^{m/2}(t)\|\le C_m.
\end{equation}
We use the following identity:
\begin{equation*}
H^{-\frac{1}{2}}(s) U_\tau (s,t) H^{\frac{1}{2}}(t) U_{\tau} (t,s) = 
1 + \int_s^t dr H^{-\frac{1}{2}}
(s) \frac{d}{dr} \left( U_\tau (s,r) H^{\frac{1}{2}}(r) U_{\tau} (r,s)\right) .
\end{equation*}
Since $\partial_r H^{1/2} (r) = i g(r) [ H^{1/2} (r) , \Lambda_1 ]$ we can 
multiply both sides by $U_{\tau} (s,t)$ from the right to get 
\begin{multline}\label{iden}
H^{-1/2}(s) U_\tau(s,t) H^{1/2}(t) \ = \ U_\tau(s,t) \\ + \int_s^t\di r \, i g(r)H(s)^{-1/2}U_\tau(s,r)[H(r)^{1/2},\Lambda_1]U_\tau(r,t)\,.
\end{multline}
Since $\|[H^{1/2}(r),\Lambda_1]\|<\infty$ (see Eq. (\ref{cob}) below
with $k=0$), last equation proves the proposition for $m=1$. For larger
values of $m$ we use induction. Eq. (\ref{iden}) implies that 
\begin{multline}
H^{-m/2}(s)U_\tau(s,t)H^{m/2}(t) \ = \
H^{-(m-1)/2}(s)U_\tau(s,t)H^{(m-1)/2}(t) \\ + i \int_s^t\di r\dot
g(r)H^{-m/2}(s)U_\tau(s,r)[H(r)^{1/2},\Lambda_1]
U_\tau(r,t)H^{(m-1)/2}(t)\,.\nonumber
\end{multline}
The first term is bounded by the induction hypothesis. The second one
is also bounded by the induction hypothesis and because of the bound:
\begin{equation}\label{cob}
\|H^{-k/2}(r)[H^{1/2}(r),\Lambda_1]H^{k/2}(r)\|\le C_k\,.
\end{equation}
To prove the last relation we use the integral representation
\begin{equation}
\sqrt H \ = \ \frac{2}{\pi} \int_0^\infty \di x \frac{H}{x^2+H}\, ,
\end{equation}
which implies that (using the concise notation $H \equiv H(r)$):
\begin{equation}
H^{-k/2}[H^{1/2},\Lambda_1]H^{k/2} = \frac{2}{\pi} \int_0^\infty \di x \, x^2 \, 
\frac{H^{-k/2}}{x^2+H} [H,\Lambda_1] \frac{H^{k/2}}{x^2+H} \, .
\end{equation}
Now we commute one of the resolvent $(x^2 + H)^{-1}$ through the commutator $[H,
\Lambda_1]$ and we find 
\begin{equation}\label{eq:com}
\begin{split}
H^{-k/2}[H^{1/2},\Lambda_1]H^{k/2} = \; &\frac{2}{\pi} \int_0^\infty \di x \, x^2 \, 
\frac{H^{-k/2}}{(x^2+H)^2} [H,\Lambda_1] H^{k/2} \\ &+ 
\frac{2}{\pi} \int_0^\infty \di x \, x^2 \, 
\frac{H^{-k/2}}{(x^2+H)^2} [H, [H,\Lambda_1]] \frac{H^{k/2}}{x^2 +H}.
\end{split}
\end{equation}
The first term on the r.h.s. of the last equation equals $H^{-(k+1)/2} [H,
\Lambda_1] H^{k/2}$ and is bounded, by Proposition \ref{prop:space}, a) - 
because $[H,\Lambda_1] = 1/2 ({\bf p_A} \cdot {\bf \nabla} \Lambda_1 + {\bf \nabla} \Lambda_1
\cdot {\bf p_A})$. On
the other hand the 
norm of the second term on the r.h.s. of (\ref{eq:com}) is bounded by 
\begin{equation}
\frac{2}{\pi} \int_0^\infty \di x \, \| \frac{x^2}{x^2 +H}\| \, \| \frac{H}{x^2
+ H} \| \, \|H^{-k/2 -1} [H, [H,\Lambda_1]] H^{k/2} \| \, \| \frac{1}{x^2 +H} \|
\end{equation}
which is finite, because $\| x^2 (x^2 +H)^{-1}\|$ and $\| H (x^2
+ H)^{-1} \|$ are bounded by 1, and $\| (x^2 +H)^{-1} \| \leq (C + x^2)^{-1}$
(this term ensures the convergence of the integral), and by 
Proposition \ref{prop:space}, $ \|H^{-k/2 -1}
[H, [H,\Lambda_1]] H^{k/2} \| \leq C_k$ (here we capitalize on the fact that 
$[H, [H,\Lambda_1]]$
is quadratic in ${\bf p_A}$). This establishes (\ref{cob}). 
\end{proof}

\begin{proof}[Proof of Proposition \ref{prop:speed}]
We only consider the case $t_1 =t$ and $t_2 =0$: if $t_2 \neq 0$ 
the proof is identical. 
We proceed by induction over $n$. If $n=0$ the claim follows by Lemma
\ref{lm:energy}. Now we take some $N\in \N/2$, we assume 
the proposition holds true for all $n=0, 1/2,1,\dots, N-1/2$ and for all 
$m\in \N/2$, and we prove it for $n=N$. To this end we use induction over $m$.
First of all we have to prove the proposition for $n=N$ and $m=0$. In the
following we denote $H_t = H_{\lambda} (t)$. We have
\begin{equation}\label{eq:m=0,0}
\begin{split}
\| \axi^N &U(t,0) H_0^{-N} \ph \|^2 = \sprod{\ph}{H_0^{-N} U(0,t) \axi^{2N} U(t,0)
H_0^{-N} \ph}\\ =\; &\sprod{\ph}{H_0^{-N} \axi^{2N} H_0^{-N} \ph} 
+ \int_0^t ds \sprod{\ph}{H_0^{-N} U(0,s) [iH(s), \axi^{2N}] U(s,0) H_0^{-N}\ph}
\\ 
=\; &\| \axi^N H_0^{-N} \ph \|^2 \\
&+\int_0^t ds \sprod{\ph}{H_0^{-N} U(0,s) \left\{ 2N \axi^{2N-1} \frac{x_i}{\axi}
\, p_{A,i} + h.c \right\} U(s,0)H_0^{-N} \ph}.
\end{split}
\end{equation}
The first term on the r.h.s. of the last equation can be estimated as  
\begin{equation}\label{eq:m=0,0b} 
\|\axi^N
H_0^{-N} \ph \|\leq \| \axi^N H_0^{-N} \axi^{-N} H_0^{N}\| \| H_0^{-N} \axi^{N} \ph
\| \leq C \| \axi^{N} \ph \|, 
\end{equation}  
by Proposition \ref{prop:space}. Now we note that
\begin{equation}\label{eq:m=0,1}
\begin{split}
\langle \ph , &H_0^{-N} U(0,s) \left\{ 2N \axi^{2N-1} \frac{x_i}{\axi} \, p_{A,i} 
+ h.c \right\} U(s,0)H_0^{-N} \ph \rangle \\ 
&\leq C_1 \sum_{i=1,2} 
\|\axi^{N-1/2} U(s,0) H_0^{-N} \ph \| \| \axi^{N-1/2} p_{A,i} U(s,0) H_0^{-N} \ph
\|.
\end{split}  
\end{equation}
The first factor on the r.h.s. can be bounded using the induction assumption
with $n=N-1/2$ and $m=0$. In the second factor, on the other hand, we commute $p_{A,i}$ to the left. We get 
\begin{equation}\label{eq:m=0,2}
\begin{split}
\| \axi^{N-1/2} p_{A,i} U(s,0) H_0^{-N} \ph \| \leq \; &\| p_{A,i} \axi^{N-1/2} U(s,0)
H_0^{-N} \ph \| \\ &+ (n-1/2) \| \axi^{N-3/2} U(s,0) H_0^{-N} \ph \|
\end{split}
\end{equation}
To handle the first term on the r.h.s. of the last equation we write 
\begin{equation}
\begin{split}
\| p_{A,i} \axi^{N-1/2} U(s,0) H_0^{-N} \ph \| \leq \; &\|  p_{A,i} H_s^{-1/2} \| \|
H_s^{1/2} \axi^{N-1/2} H_s^{-1/2} \axi^{-N+1/2} \| \\ &\times \| \axi^{N-1/2} H_s^{1/2} U(s,0)
\ph \|,
\end{split}
\end{equation}
where the factor $H_s^{1/2} \axi^{N-1/2} H_s^{-1/2} \axi^{-N+1/2}$ is bounded,
uniformly in $s$, by Proposition \ref{prop:space}, part b). 
Now we insert the last equation into (\ref{eq:m=0,2}) and we substitute the 
result into (\ref{eq:m=0,1}). Then we use the induction assumption 
for $n=N-1/2$ and $m=1/2$ and for $n=N-3/2$ and $m=0$ (in order to bound the
contribution of the second term on the r.h.s. of (\ref{eq:m=0,2})), and we get 
\begin{equation}
\begin{split}
\langle \ph , H_0^{-N} &U(0,s) \left\{ 2N \axi^{2N-1} \frac{x_i}{\axi}
 \, p_{A,i} + h.c \right\} U(s,0)H_0^{-N} \ph \rangle \\ &\leq C (s^{N-1/2} +1 )^2 \|
\axi^{N-1/2} \ph \|^2
\end{split}
\end{equation}
Inserting this equation and (\ref{eq:m=0,0b}) in the r.h.s. of (\ref{eq:m=0,0})
and performing the integration over $s$ we find
\begin{equation}
\| \axi^N U(t,0) H_0^{-N} \ph \|^2 \leq C (t^{2N}+1) \| \axi^N \ph \|^2
\end{equation}
which proves the proposition for $n=N$ and $m=0$.

Now we assume that the proposition holds true for $n\leq N-1/2$ and all $m\in
\N/2$ and also for $n=N$ and $\N/2 \ni m \leq M-1/2$, for some $M \in \N/2$, and
we verify it for $n=N$ and $m=M$. We compute
\begin{equation}\label{eq:m>0,1}
\begin{split}
\| \axi^N &H_t^M U(t,0) H_0^{-N-M} \ph \|^2 \\ =\; & \sprod{\ph}{H_0^{-N-M} U(0,t)
H_t^M \axi^{2N} H_t^M U(t,0) H_0^{-N-M} \ph} \\ = \; &\sprod{\ph}{H_0^{-N}\axi^{2N}
H_0^{-N} \ph} \\ &+ \int_0^t ds \frac{d}{ds} \, 
\sprod{\ph}{H_0^{-N-M} U(0,s) H_s^M \axi^{2N} H_s^M U(s,0) H_0^{-N-M} \ph} \\
= \; &\sprod{\ph}{H_0^{-N}\axi^{2N}
H_0^{-N} \ph} \\ &+ \int_0^t ds \, 
\sprod{\ph}{H_0^{-N-M} U(0,s) H_s^M [iH(s) , \axi^{2N}] H_s^M U(s,0) H_0^{-N-M}
\ph} \\
&+\int_0^t ds \, \sprod{\ph}{H_0^{-N-M} U(0,s) \left\{ (\frac{d}{ds} H_s^M )
\axi^{2N} H_s^M  + h.c \right\} U(s,0) H_0^{-N-M} \ph}. 
\end{split}
\end{equation}
The first term on the r.h.s. of the last equation can be bounded, using
Proposition \ref{prop:space}, part b), by
\begin{equation}\label{eq:m>0,1c}
\begin{split}
\sprod{\ph}{H_0^{-N}\axi^{2N}
H_0^{-N} \ph} &= \| \axi^N H_0^{-N} \ph \|^2 \\ &\leq \|\axi^N H_0^{-N} \axi^{-N}
H_0^{N}\|^2 \| H_0^{-N} \|^2 \| \axi^N  \ph \|^2 \\ &\leq C \| \axi^N  \ph \|^2 .
\end{split}
\end{equation}
The second term on the r.h.s. of (\ref{eq:m>0,1}) can be controlled in the same
way as we did with the second term on the r.h.s. of (\ref{eq:m=0,0}). We find,
applying the induction assumption for $n=N-1/2$ and $m=M$
\begin{equation}\label{eq:m>0,1b}
\begin{split}
\int_0^t ds \, \langle \ph , H_0^{-N-M} U(0,s) H_s^M [iH(s) , \axi^{2N}] &H_s^M
U(s,0) H_0^{-N-M}\ph \rangle \\ &\leq C (t^{2N}+1) \| \axi^{N-1/2} \ph \|^2.  
\end{split}
\end{equation}
We consider now the third term on the r.h.s. of (\ref{eq:m>0,1}), and we first
assume that $M \in \N$. Then we have 
\begin{equation}
\frac{d}{ds} \, H_s^{M} = \sum_{j=1}^M H_s^{j-1} \dot{H_{\lambda}} (s) H_s^{m-j}
\end{equation}
where $\dot{H_{\lambda}} (s) = 1/\tau^2 \dot{g} (s/\tau) \Lambda_1$. Note here 
that $\dot{H_{\lambda}} (s) =0$ if $s > \tau$. From last equation it follows that
\begin{equation}\label{eq:m>0,2} 
(\frac{d}{ds} H_s^M ) \axi^{2N} H_s^M + h.c. = \frac{\dot{g} (s/\tau)}{\tau^2} 
\sum_{j=1}^M \left( H_s^{j-1} \Lambda_1 H_s^{M-j} \axi^{2N} H_s^M + h.c. \right). 
\end{equation}
To handle this term we note that, for any $j =1,2, \dots M$, there is a constant
$D <\infty$ such that  
\begin{equation}\label{eq:m>0,3}
 H_s^{j-1} \Lambda_1 H_s^{M-j} \axi^{2N} H_s^M + h.c. \leq D H_s^{M-1/2}
 \axi^{2N} H_s^{M-1/2}. 
\end{equation}
To prove the last equation we set $A=H_s^{M-1/2}
\axi^{2N} H_s^{M-1/2}$. First of all we note that $A \geq 0$ and that \[ \| 
\frac{1}{\sqrt{A}} H_s^{M/2-1/4} \axi^N H_s^{M/2-1/4} \| < \infty .\] In fact 
for any $\ph \in \Hi$ we have 
\begin{equation}
\begin{split}
\| \frac{1}{\sqrt{A}} &H_s^{M/2-1/4} \axi^N H_s^{M/2-1/4} \ph\|^2 \\ 
&=\sprod{\ph}{
H_s^{M/2-1/4} \axi^N H_s^{M/2-1/4} \, A^{-1} \, H_s^{M/2-1/4} \axi^N
H_s^{M/2-1/4}\ph} \\ &= \sprod{\ph}{
H_s^{M/2-1/4} \axi^N H_s^{-M/2+1/4} \axi^{2N} H_s^{-M/2+1/4} \axi^N
H_s^{M/2-1/4}\ph} \\
&\leq \| H_s^{M/2-1/4} \axi^N H_s^{-M/2+1/4} \axi^{N} \|^2 \| \ph \|^2 \\
&\leq C \| \ph \|^2 ,
\end{split}
\end{equation}
where we used Proposition \ref{prop:space}, part b). 
Now, if we denote by $B$ the operator on 
the l.h.s. of (\ref{eq:m>0,3}) we have  
\begin{equation}
B= \sqrt{A} \frac{1}{\sqrt{A}} B \frac{1}{\sqrt{A}} \sqrt{A}
\end{equation}
and (\ref{eq:m>0,3}) follows if we prove that $ A^{-1/2} B A^{-1/2}$ is bounded.
Since we know that $A^{-1/2} H_s^{M/2-1/4} \axi^N H_s^{M/2-1/4}$ is bounded, it
is enough to prove the boundedness of
\begin{equation}
\begin{split}
&H_s^{-M/2+1/4} \axi^{-N} H_s^{-M/2+1/4} B H_s^{-M/2+1/4} \axi^{-N} H_s^{-M/2+1/4}
 \\ &= H_s^{-M/2+1/4} \axi^{-N} H_s^{-M/2+j-3/4} \Lambda_1 H_s^{M-j} 
\axi^{2N} H_s^{M/2+1/4} \axi^{-N} H^{-M/2+1/4} \\ 
&= C_1 \axi^{-N} H_s^{-M+ j -1/2} \Lambda_1 H_s^{M-j} \axi^{2N} H_s^{1/2} \axi^{-N}
C_2,
\end{split}
\end{equation}
where, by Proposition \ref{prop:space}, 
the operators $C_1 =   H_s^{-M/2+1/4} \axi^{-N} H_s^{M/2-1/4} \axi^{N}$ and 
$C_2 = \axi^{N} H^{M/2-1/4} \axi^{-N} H^{-M/2+1/4}$ are bounded. 
Using part b) of that statement 
we can exchange the operators
$\axi^{-N}$ with the powers of the Hamiltonian $H_s$ once again. The operator on
the r.h.s. of the last equation can thus be written as
\begin{equation*}
\begin{split}
\tilde{C}_1 H_s^{-M+ j -1/2} \axi^{-N} \Lambda_1 H_s^{M-j} &\axi^{N}
H_s^{1/2} \tilde{C}_2 = \tilde{C}_1 H_s^{-M+ j -1/2} \Lambda_1 H_s^{M -j+1/2}
\\ &\times H_s^{-M+j-1/2}\axi^{-N} H_s^{M-j} \axi^{N} H_s^{1/2} \tilde{C}_2, 
\end{split}
\end{equation*}
where the operators $\tilde{C}_1$ and $\tilde{C}_2$ are bounded. Because of
Proposition \ref{prop:space}, part a), the operator $H_s^{-M+ j -1/2} \Lambda_1 H_s^{M -j+1/2}$
is bounded. So, if we exchange the operators $H_s^{-M+j-1/2}$ and $\axi^{-N}$,
using Proposition \ref{prop:space}, part b), the operators on the r.h.s. of the last equation
becomes
\begin{equation}
\tilde{\tilde{C}}_1 \axi^{-N} H_s^{-1/2} \axi^{N} H_s^{1/2} \tilde{C}_2,
\end{equation}
for some bounded operator $\tilde{\tilde{C}}_1$. But now, because of Proposition
\ref{prop:space}, 
also the operator $ \axi^{-N} H_s^{-1/2} \axi^{N} H_s^{1/2}$ is bounded. This
establishes that the
operator $A^{-1/2} B A^{-1/2}$ is bounded and completes the proof of
Eq. (\ref{eq:m>0,3}). 

Plugging Eq. (\ref{eq:m>0,3}) into the r.h.s. of
(\ref{eq:m>0,2}) we observe that the third term on the r.h.s. of (\ref{eq:m>0,1})
is bounded by
\begin{equation}\label{eq:m>0,4}
\begin{split}
&\int_0^t ds \, \sprod{\ph}{H_0^{-N-M} U(0,s) \left\{ (\frac{d}{ds} H_s^M )
\axi^{2N} H_s^M  + h.c \right\} U(s,0) H_0^{-N-M} \ph}  \\ &\leq 
\frac{C}{\tau^2}
\int_0^t ds \chi (s \leq \tau) \sprod{\ph}{H_0^{-N-M} U(0,s) H_s^{M-1/2} \axi^{2N} H_s^{M-1/2} U(s,0) H_0^{-N-M}
\ph} \\ &\leq \frac{C}{\tau^2} \int_0^{\min (t , \tau)} (s^{2N} +1) \| \axi^N \ph
\|^2 \leq \frac{C}{\tau^2} \left( \min (t , \tau)^{2N+1} + 1 \right) \| \axi^N \ph
\|^2 \\ &\leq C (t^{2N-1} +1) \|\axi^N \ph \|^2 \, ,
\end{split}
\end{equation}
where we used the induction assumption for $n=N$ and $m=M-1/2$. A similar result
can also be proved if $M \in \N/2$ is not an integer. Inserting (\ref{eq:m>0,4}), (\ref{eq:m>0,1b}) and (\ref{eq:m>0,1c}) into
(\ref{eq:m>0,1}) we finally find that 
\begin{equation}
\| \axi^N H_t^M U(t,0) H_0^{-N-M} \ph \|^2 \leq C (t^{2N}+1) \| \axi^N \ph \|^2
\, .
\end{equation}
\end{proof}
\noindent{\em Acknowledgments.} This work is supported in part by the NSF Grant DMS-9983190.
A.E. is grateful to Yosi Avron for the hospitality at Technion, where part of this work
was done. 

\end{document}